\newcommand{\beq}{\begin{equation}}
\newcommand{\eeq}{\end{equation}}
\newcommand{\bea}{\begin{eqnarray}}
\newcommand{\eea}{\end{eqnarray}}

\newcommand{\gsim}{\lower.7ex\hbox{$\;\stackrel{\textstyle>}{\sim}\;$}}
\newcommand{\lsim}{\lower.7ex\hbox{$\;\stackrel{\textstyle<}{\sim}\;$}}

\newcommand{\mrm}{\mathrm}

\documentclass[aps,prev,twocolumn,preprintnumbers,floatfix,nofootinbib]{revtex4-1}
\usepackage{graphicx}
\usepackage{epstopdf}
\usepackage{mathrsfs}
\usepackage{amssymb}
\usepackage{verbatim}
\usepackage{color}
\usepackage{multirow}
\usepackage{amsmath}
\usepackage{subfig}
 \usepackage{slashed} 
\usepackage{float}

\def\stacksymbols #1#2#3#4{\def\theguybelow{#2}
    \def\vp{\lower#3pt}
    \def\sp{\baselineskip0pt\lineskip#4pt}
    \mathrel{\mathpalette\intermediary#1}}

\def\intermediary#1#2{\vp\vbox{\sp
     \everycr={}\tabskip0pt
     \halign{$\mathsurround0pt#1\hfil##\hfil$\crcr#2\crcr
              \theguybelow\crcr}}}

\def\be{\begin{equation}}
\def\ee{\end{equation}}
\def\bea{\begin{eqnarray}}
\def\eea{\end{eqnarray}}

\def\sp{\;\;\;,\;\;\;}

\def\mrm{\mathrm}

\def\lsim{\raise0.3ex\hbox{$\;<$\kern-0.75em\raise-1.1ex\hbox{$\sim\;$}}}
\def\gsim{\raise0.3ex\hbox{$\;>$\kern-0.75em\raise-1.1ex\hbox{$\sim\;$}}}

\def\inbar{\,\vrule height1.5ex width.4pt depth0pt}

\def\IC{\relax\hbox{$\inbar\kern-.3em{\rm C}$}}
\def\IQ{\relax\hbox{$\inbar\kern-.3em{\rm Q}$}}
\def\IR{\relax{\rm I\kern-.18em R}}
 \font\cmss=cmss10 \font\cmsss=cmss10 at 7pt
\def\IZ{\relax\ifmmode\mathchoice
 {\hbox{\cmss Z\kern-.4em Z}}{\hbox{\cmss Z\kern-.4em Z}}
 {\lower.9pt\hbox{\cmsss Z\kern-.4em Z}}
 {\lower1.2pt\hbox{\cmsss Z\kern-.4em Z}}\else{\cmss Z\kern-.4em Z}\fi}

\def\comment#1{}
\def\to{\rightarrow}

\def\u1x{U(1)_X}
\newcommand{\nc}{\newcommand}
\nc{\LL}{L}
\nc{\vv}{\tilde{v}}
\nc{\ccdot}{\!\cdot\!}
\nc{\gsm}{G_{SM}}
\nc{\vfive}{\mathbf{5}\oplus\mathbf{\overline{5}}}
\nc{\vten}{\mathbf{10}\oplus\mathbf{\overline{10}}}
\nc{\zhol}{Z^{\rm hol}}
\nc{\xfb}{\,{\rm fb}}

\begin{document}

%
%

\preprint{CPHT-RR053.102017}
\preprint{LPT--Orsay 17-41}
\preprint{UMN--TH--3704/17}
\preprint{FTPI--MINN--17/20}

\vspace*{1mm}

\title{Inflation and High-Scale Supersymmetry with an EeV Gravitino}

\author{Emilian Dudas$^{a}$}
\email{Emilian.Dudas@cpht.polytechnique.fr}
\author{Tony Gherghetta$^{b}$}
\email{tgher@umn.edu}
\author{Yann Mambrini$^{c}$}
\email{yann.mambrini@th.u-psud.fr}
\author{Keith A. Olive$^{b,d}$}
\email{olive@physics.umn.edu}

\vspace{0.1cm}
\affiliation{
${}^a$CPhT, Ecole Polytechnique, 91128 Palaiseau Cedex, France \\
${}^b$School of Physics and Astronomy, University of Minnesota, Minneapolis, MN 55455, USA\\
${}^c$Laboratoire de Physique Th\'eorique 
Universit\'e Paris-Sud, F-91405 Orsay, France\\
${}^d$William I.~Fine Theoretical Physics Institute, University of Minnesota, Minneapolis, MN 55455, USA
}

\begin{abstract} 

We consider inflation and supersymmetry breaking in the context of a minimal model of 
supersymmetry in which the only ``low" energy remnant of supersymmetry is the gravitino 
with a mass of order an EeV. In this theory, the supersymmetry breaking scale
is above the inflaton mass, $m \simeq 3 \times 10^{13}$ GeV, as are all 
sfermion and gaugino masses. In particular, for a no-scale formulation of Starobinsky-like inflation
using the volume modulus $T$, we show that inflation can be accommodated even when the supersymmetry breaking
scale is very large. Reheating is driven through a gravitational coupling to the
two Higgs doublets and is enhanced by the large 
$\mu$-parameter. This leads to gravitino cold dark matter where the mass is constrained to be in the range $0.1~{\rm EeV} \lesssim m_{3/2} \lesssim 1000~{\rm EeV}$.
\\

\end{abstract}

\maketitle


\maketitle


\setcounter{equation}{0}





\section{Introduction}
While the Higgs boson was discovered at the LHC \cite{lhch,125} 
and is consistent with predictions of low energy supersymmetry (SUSY),
\cite{mh,mh2loop,mhiggsAEC,susycompare,ENOS}, so far supersymmetry has not 
been  seen experimentally \cite{nosusy}. Whether supersymmetry is waiting 
around the corner, or  is  broken at some high scale (intermediate  or above)
is currently unknown. If indeed supersymmetry is broken above the inflationary
scale, it may well be that the only remnant of supersymmetry at low energies
is the gravitino which may yet play the role of dark matter \cite{bcdm,DMO}\footnote{In this case supersymmetry is 
nonlinearly realized at lower energies \cite{nonlinear}.}.

Below the Planck and grand unified theory (GUT) scales, it would appear that there is an
intermediate scale (between the GUT scale and electroweak scale) associated with  inflation. For the sake of definiteness, let us consider the Starobinsky model of inflation as an example \cite{Staro,MC,Staro2}. The inflaton potential can be written as
\beq
V(t) =  \frac{3}{4} m^2 \left(1- e^{-\sqrt{\frac{2}{3}}t}\right)^2 \, ,
\label{inflpot}
\eeq
where $t$ is the canonically normalized inflaton field.
The inflaton mass scale, $m$, can be determined by the amplitude of density fluctuations \cite{planck15},
\beq
A_s=\frac{3m^2}{8\pi^2} \sinh^4(t_*/\sqrt{6}) = 2.1 \times 10^{-9}~,
\label{As}
\eeq
where $t_* \approx 5.35$ corresponds to 55 efolds of inflation. 
Solving for $m$ in (\ref{As}), we have $m = 1.2\times 10^{-5} M_P \approx 3 \times 10^{13} $ GeV, where $M_P = 1/\sqrt{8\pi G_N}\simeq 2.4\times 10^{18}$ GeV.  

Interestingly, the mass scale around $10^{13}$ GeV, may also correspond to an intermediate 
scale gauge group whose breaking may yield a large Majorana mass for right-handed neutrinos, $M_R \simeq m$,
appropriate for the see-saw mechanism \cite{seesaw}. 

Here, we consider the possibility that the supersymmetry breaking scale is also of order the inflaton mass, $m$.  
An example is provided by
supersymmetry breaking via a Polonyi sector which is achieved
with a superpotential of the form \cite{pol}
\beq
W_{P} = {\widetilde m}^2 (Z + b)\,,
\label{wpol}
\eeq
where $Z$ is the chiral superfield responsible for breaking supersymmetry with auxiliary field component $F\equiv {\widetilde m}^2$. 
Of course if ${\widetilde m}$ in (\ref{wpol}) is of order the inflaton mass, $m$, then the masses of the entire
supersymmetric spectrum would be of order the intermediate scale and clearly out of 
reach of any accelerator search. However, the gravitino mass,
\beq
m_{3/2} = \frac{{\widetilde m}^2}{\sqrt{3}M_P}~,
\eeq
would be significantly lighter and could still provide for the dark matter in the universe \cite{bcdm,DMO}.
In fact, to avoid over production of gravitinos through the decay of the inflaton to $R$-parity = -1 matter
fields (which subsequently decay to gravitinos), it was argued \cite{DMO}, that the sparticle spectrum
should lie above the inflaton mass, thus providing a lower limit to the supersymmetry breaking scale
and hence a lower limit to the gravitino mass of  $m_{3/2} > 0.2$ EeV.
Only an EeV gravitino mass is left behind.

The EeV gravitino as a dark matter candidate is produced after inflation in the process of reheating.
The common  mechanism 
\cite{nos,ehnos,kl,ekn,Juszkiewicz:gg,mmy,Kawasaki:1994af,Moroi:1995fs,enor,Giudice:1999am,bbb,Pradler:2006qh,ps2,rs,egnop}, for 
producing a single gravitino in thermal scattering processes has a cross section which is temperature independent and scales 
as $m_{\rm SUSY}^2/M_P^2 m_{3/2}^2$, where $m_{\rm SUSY}$ is a typical sparticle mass. The rate therefore is 
roughly $\Gamma \sim T^3 m_{\rm SUSY}^2/M_P^2 m_{3/2}^2$, where we have assumed predominantly Goldstino production in the limit $m_{3/2} \ll m_{\rm SUSY}$.
If sparticle production is kinematically forbidden, single gravitino production (which must be accompanied by a massive
sparticle (gluino),  is not operative.
Instead, the rate for gravitino production during reheating is suppressed, 
as only processes which produce two gravitinos are allowed.  This cross section is temperature dependent and scales 
as $\langle \sigma v \rangle
\propto T^6/F^4$, so that the rate is roughly $\Gamma \sim T^9/F^4$.  In this case, the final gravitino abundance
scales as $n_{3/2}/n_\gamma \sim \Gamma/H \sim T^7 M_P/F^4$ evaluated at the reheating temperature, in contrast to the abundance for single gravitino production,  $n_{3/2}/n_\gamma \sim \Gamma/H \sim T m_{\rm SUSY}^2/M_P m_{3/2}^2$. For reheating temperatures of order $10^{10}$ GeV, the gravitino 
abundance matches the CMB determined cold dark matter density \cite{planck15}. 
It is also possible that the inflaton can decay to two gravitinos, but this is more model dependent and we return to this 
possibility in section III.D.

The phenomenology of this high scale supersymmetric model is simple.  The (not so) low energy spectrum
consists of the gravitino and perhaps the scalars associated with the chiral superfield, $Z$.  However,
as we discuss in section III.A, we expect that these are also hierarchically more massive than the gravitino.
As has been shown recently \cite{susyhd,ellisjr}, even with a spectrum as massive as discussed here,
a 125 GeV Higgs mass can still be attained if $\tan \beta$ is either small (close to 1) or large (above 60). 

The clear drawback of such a model is its testability. 
In fact, the model in its simplest and most minimal form predicts no
signatures in either accelerator searches, or direct and indirect searches for dark matter. 
Of course if supersymmetry is actually discovered at the LHC, 
then this model can be ruled out. In a modest extension of the model with
R-parity violation in the lepton-Higgs sector, the gravitino becomes
unstable, though still suitably long lived. The detection of very high energy neutrinos or photons at
HAWC or the Pierre Auger Observatory
would be a signature of this model. Another possible signature may 
come from the observation of non-gaussianities in the CMB due to scalars with masses
near the Hubble scale during inflation \cite{fnl}. 

The paper is organized as follows. 
In section II, we discuss exemplary inflation models with high scale supersymmetry breaking.
We focus on models based on no-scale supergravity \cite{no-scale,LN} which lead to Starobinsky-like 
potentials \cite{ENO6,KLno-scale,fkr,ENO7,ENO8,adfs,others,EGNO4,EGNO5,DW,EGNO6,egnno,king,egnno2}.
In particular, models which allow for high scale supersymmetry breaking with stabilized fields \cite{ENO6,KLno-scale,ENO7,adfs,EGNO4,DW} 
without spoiling the inflationary
properties of the potential. In section III, we discuss the phenomenological aspects of the model.
We begin in section III.A with a model for gaugino and scalar masses.  
The model utilizes the strong stabilization of the Polonyi field 
\cite{dine,Dudas:2006gr,klor,dlmmo,eioy,nataya,ADinf,ego,EGNO4} used to generate large gaugino masses.  
Scalar masses are then obtained
through threshold corrections as in gaugino mediation \cite{gaugino_mediation} or a more strongly-coupled mediation mechanism~\cite{warped,ggm}. The requirement that all sparticle masses
lie above the inflaton mass will set a constraint on the stabilization scale. 
The effects of supersymmetry breaking on inflation is then discussed in Section III.B.
In section III.C we determine the conditions under which we can obtain a Higgs mass of 125 GeV, as well as preserving the stability of the Higgs vacuum. 
The requirements for gravitino dark matter are outlined in section III.D. 
Our concluding remarks are given in section IV.

\section{Inflation and Supersymmetry breaking}

There are many ways to proceed in constructing a model of inflation
which incorporates supersymmetry breaking. While it would be an overstatement to
say that recent Planck results \cite{planck15} on the CMB spectrum parameters, $n_s$ and $r$, corresponding
to the tilt of the scalar perturbation spectrum and the scalar to tensor ratio, respectively predict Starobinsky-like 
inflation models, it is clear that these models are for now in very good agreement with Planck results. 
In particular, we will use formulations of the Starobinsky model based on no-scale supergravity. 

\subsection{No-scale supergravity and Starobinsky-like Inflation}

The K\"ahler potential in the context of no-scale supergravity can be written in the form \cite{no-scale,LN},
\beq
K \; = \; - 3 \ln \left(T + {\bar T} - \frac{1}{3} \sum_i |\phi_i|^2\right) \,,
\eeq
where $T$ is a volume modulus and the $\phi_i$ include all matter fields and 
possibly a supersymmetry breaking Polonyi-like field, $Z$ \cite{pol}.
The inflaton may be identified with either $T$ or a matter-like field, $\phi$.  
Equivalently, we may use a set of field redefinitions \cite{EKNGUT} and write
\begin{equation}
K \; = \; - 3 \ln \left(1 - \sum_i\frac{|y_i|^2}{3} \right) \, ,
\label{K21symm}
\end{equation}
where the $y_i$ include all matter fields, moduli and the inflaton.
For now, let us ignore the matter fields, and concentrate on a
two-field model. 
There are at least two independent families \cite{ENO7} of superpotentials which
lead to Starobinsky-like inflation. 
In the first, $T$ is a modulus and $\phi$ is the inflaton with a Wess-Zumino (WZ)
superpotential written as \cite{ENO6}
\beq\label{WZ_staro}
W=m\left(\frac{\phi^2}{2}-\frac{\phi^3}{3\sqrt{3}}\right) \qquad {\rm WZ} \, ,
\eeq
or in the symmetric basis
\begin{equation}
W \; = \; m \left[ \frac{y_1^2}{2} \left(1+\frac{y_2}{\sqrt{3}} \right) - \frac{y_1^3}{3 \sqrt{3}} \right] \qquad {\rm WZ} \, ,
\label{W1}
\end{equation}
which is a WZ model for the inflaton $y_1$ with an interaction term $y_1^2 y_2$.
In both bases, when $\phi$ ($y_1$) is redefined to a field $x$ with a canonical kinetic term,
the potential is exactly of the form of the  Starobinsky potential (\ref{inflpot}) (with $t$ identified as $x$), assuming that some dynamics stabilizes and 
fixes $T$ ($y_2$): $\langle T \rangle = \langle T^* \rangle = 1/2$ ($\langle y_2 \rangle= 0$).
One way to accomplish this is by adding a quartic term in the K\"ahler potential \cite{EKN3,ENO7}.

The second family of models first formulated as an $R^2$ extension to supergravity by Cecotti \cite{Cecotti} can be written as 
\beq\label{tph_w}
W=\sqrt{3}m\phi(T-1/2) \qquad {\rm C} \, ,
\eeq
or
\begin{equation}
W \; = \; m y_1 y_2 (1 + y_2/\sqrt{3} ) \qquad {\rm C} \, ,
\label{Ex3}
\end{equation}
In this case, the inflaton is associated with $T$ ($y_2$) and it must be assumed that $\phi$ ($y_1$) is stabilized at the origin \cite{KLno-scale,ENO7}.
Again, when $T$ ($y_2$) is normalized to give a proper kinetic term, we get the Starobinsky potential shown in Eq. (\ref{inflpot}) \cite{KLno-scale}.

In either case (WZ or C), the mass parameter $m$ is related to the inflaton mass,
and is set by the amplitude of density fluctuations measured in the CMB through Eq. (\ref{As}).

\subsection{Effects of supersymmetry breaking on Inflation}

Supersymmetry breaking can be accomplished in various ways, 
but in many of these, there are constraints on the SUSY breaking scale
due to its effect on the inflationary potential.
In general, SUSY breaking perturbs the potential, but these effects may be small,
if the supersymmetry breaking scale, ${\widetilde m} \ll m$. Indeed, this was one of the initial motivations
behind supersymmetric formulations of inflation \cite{ENOT}. 

The simplest possibility we can consider is adding a constant, $w_0$ to the superpotential.
In most low energy models of SUSY phenomenology, 
we would relate $w_0$ to the weak scale through $w_0 = {\tilde m} M_P^2$ 
and the gravitino mass is just 
\beq
m_{3/2} = \frac{w_0}{(T+\bar{T})^{3/2}} = {\tilde m}\,,
\eeq
with $T+\bar{T} = 1$, in Planck units. 
However, in this case low energy SUSY breaking parameters such as 
soft scalar masses, $m_0$, trilinear $A$-terms and the bi-linear $B_0$ are all proportional to $m_{3/2}$
($m_0 = 0$ for untwisted matter fields) \cite{EGNO4,lnr}.
The gaugino mass in this case is 
\beq
M_{1/2} = \left| \frac{1}{2} e^{G/2}\frac{\bar{f}_{T}}{{\rm Re} f} (G^{-1})^T_T G^T \right| =  \left| \frac{1}{2} w_0 \frac{\bar{f}_{T}}{{\rm Re}\,f}  \right| \,,
\eeq
where $f_{\alpha\beta} = f  \delta_{\alpha\beta} $ is the gauge kinetic function. For $T+\bar{T} = 1$, 
it is unlikely that we get a hierarchy $m_{3/2} \ll M_{1/2}$.
Moreover, trying to relate the supersymmetry breaking scale to the inflationary scale in this case is rather arbitrary, as we can set $w_0$ to be 
either $m M_P^2$, or $m^2 M_P$, or $m^3$, giving 
$m_{3/2} = m, m^2/M_P \approx 0.4\,{\rm EeV}$, or $m^3/M_P^2 \approx 5$ TeV (though the latter may be of phenomenological interest at the LHC). 

In \cite{king}, a linear term $a^2 \phi$ for the inflaton in the WZ model given in (\ref{WZ_staro}) was proposed, making the 
association between the inflaton and Polonyi field. For small $a$, the theory works quite well,
and thus predicts a small (weak scale) gravitino mass.  The inflationary
capability of the theory breaks down when $a \gtrsim 5 \times 10^{-5}$ corresponding 
to an upper limit on the gravitino mass of $m_{3/2} = a^4/2m \lesssim 10^6$ GeV.

\begin{figure}[h!]
\centering
	\scalebox{0.5}{\includegraphics{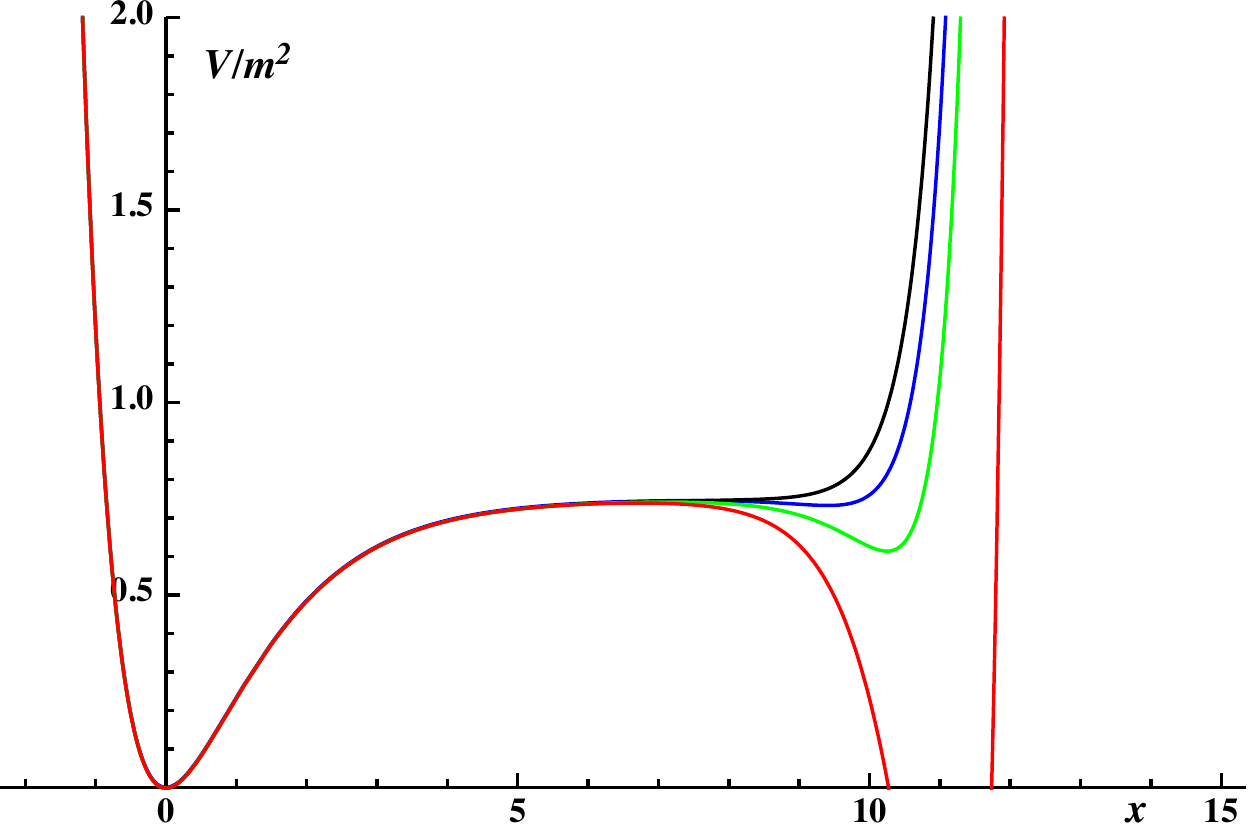}} 
	\caption{\it Projections of the effective inflationary potential for the model (\ref{WZ_staro}) with the
	Polonyi sector ((\ref{mu_pol_2}) and (\ref{K_pol})),
	for $p=3$. Here $\langle T\rangle =1/2$, $\langle z\rangle \simeq \Lambda_z^2/\sqrt{12}$ and $b\simeq 1/\sqrt{3}$, and
	we use the nominal value $\Lambda_z = 10^{-2}$. Shown is the potential for different choices of ${\tilde m}^2/m=10^{-8}, 10^{-7}, 2\times10^{-7}, 5\times10^{-7}$ (in Planck units) in black, blue, green, and red. 
	} 
	\label{pol_lam2}
\end{figure}

Next we can consider adding a strongly stabilized (twisted) Polonyi field to the WZ model,
\beq\label{mu_pol_2}
w_0 \rightarrow {\tilde m}^2(z+b)(T+1/2)^{p}\,,
\eeq
with 
\beq\label{K_pol}
K \supset z\bar{z}-\frac{(z\bar{z})^2}{\Lambda_{z}^2}\,.
\eeq
The factor  $(T+1/2)^p$ is needed to avoid a deSitter vacuum with weak scale energy density and we
will take $p=3$ as an example here \cite{EGNO4}. 
Choosing $b \simeq 1/\sqrt{3}$ gives a minimum with zero 
vacuum energy at $\langle z\rangle \simeq \Lambda_z^2/\sqrt{12}$.
The mass of the Polonyi field is now hierarchically larger than the gravitino mass
\beq
m_z^2 = \frac{12 m_{3/2}^2 M_P^2}{\Lambda_z^2}\,.
\label{zmass}
\eeq
Once again, for ${\tilde m}^2 \ll m M_P$, this works quite well so long as $\Lambda_z$ 
is not too small ($\Lambda_z \gtrsim 2({\tilde m}^2/mM_P)^{.3}$).
Increasing ${\tilde m}$, leads to the formation of a new minimum at large field values
(of the canonically normalized inflaton), which quickly becomes the global minimum.
In Fig.~\ref{pol_lam2}, we show the potential for fixed $p=3$, $\Lambda_z = 10^{-2}$,  and several values of ${\tilde m}$.
For ${\tilde m}^2/m < 10^{-8} M_P$, the potential is indistinguishable from that shown as $10^{-8}$. For $m \approx 10^{-5} M_P$, we see
that this model works fine for weak scale supersymmetry breaking, and for scales as large as 
${\tilde m}^2/M_P \lesssim 10^{-12} M_P \sim 3$ PeV, corresponding to a gravitino mass of $m_{3/2} \simeq
{\tilde m}^2/\sqrt{3}M_P \lesssim 1.7$ PeV. In particular, $m_{3/2} = 0.2$ EeV would correspond to ${\tilde m}^2/m \approx
10^{-5}M_P$ which would badly spoil the inflationary potential.

While the WZ models are perfectly acceptable for low scale supersymmetry breaking, our objective here is 
high scale breaking and thus we 
turn our attention to the case C, for the superpotential given by (\ref{tph_w}). 
To achieve supersymmetry breaking and generate a finite gravitino mass, 
we can again add a constant, $w_0$ to the superpotential.  In this case, 
if $w_0 \ll m$, the minimum is shifted slightly
to \cite{EGNO4}
\beq\label{Tinfmu}
\langle T\rangle=\frac{1}{2}-\frac{w_0^2}{m^2} \ , \quad \langle \phi\rangle = \sqrt{3}\frac{w_0}{m}\ ,
\eeq
but the vacuum energy density is necessarily negative, $V_0=-3\langle e^{G}\rangle = -3m^2w_0^2/(m^2-3w_0^2)<0$.

However, adding an untwisted Polonyi field, so that the K\"ahler potential becomes
\beq
K \; = \; - 3 \ln \left(T + {\bar T} - \frac{1}{3} \sum_i |\phi_i|^2 -\frac{1}{3}  |z|^2 + \frac{|z|^4}{\Lambda_z^2} \right) \,,
\label{Kpol2}
\eeq
with the superpotential given in  (\ref{mu_pol_2}) with $p=0$ leaves the Starobinsky potential
(now a function of $T$) unchanged, save for a shift in the minimum to 
\begin{eqnarray}
\langle T\rangle \simeq \frac{1}{2} + \frac{1}{3}\left(\frac{{\tilde m}^2}{mM_P}\right)^2\ , \ \ \langle\phi\rangle\simeq \frac{{\tilde m}^2}{m}\ , \nonumber \\
\langle z \rangle\simeq \frac{\Lambda_z^2}{6\sqrt{3}}\ , \ \ b \simeq\frac{1}{\sqrt{3}}\left(1-\frac{1}{6}\left(\frac{{\tilde m}^2}{mM_P}\right)^2\right) \, ,
\label{shift}
\end{eqnarray}
when ${\tilde m}^2/(mM_P), \Lambda_z/M_P \ll 1$. The mass of $z$ is $\sqrt{3}$ times larger than the twisted Polonyi mass given in Eq. (\ref{zmass}).

Alternatively, one can add a twisted Polonyi field with K\"ahler potential
\beq
K \; = \; - 3 \ln \left(T + {\bar T} - \frac{1}{3} \sum_i |\phi_i|^2\right) + |z|^2 - \frac{|z|^4}{\Lambda_z^2}  \,,
\label{Kpol3}
\eeq
and the same superpotential (\ref{tph_w}).
This also leaves the Starobinsky potential
unchanged, with a similar shift in the minimum to \cite{EGNO4}
\begin{eqnarray}
\langle T\rangle \simeq \frac{1}{2} + \frac{2}{3}\left(\frac{{\tilde m}^2}{mM_P}\right)^2\ , \ \ \langle \phi\rangle\simeq \frac{{\tilde m}^2}{m}\ , \nonumber \\
\langle z \rangle \simeq \frac{\Lambda_z^2}{2\sqrt{3}}\ , \ \ b \simeq\frac{1}{\sqrt{3}}\left(1-\frac{1}{2}\left(\frac{{\tilde m}^2}{mM_P}\right)^2\right) \, ,
\label{shift2}
\end{eqnarray}

\begin{figure}[h!]
\centering
	\scalebox{0.5}{\includegraphics{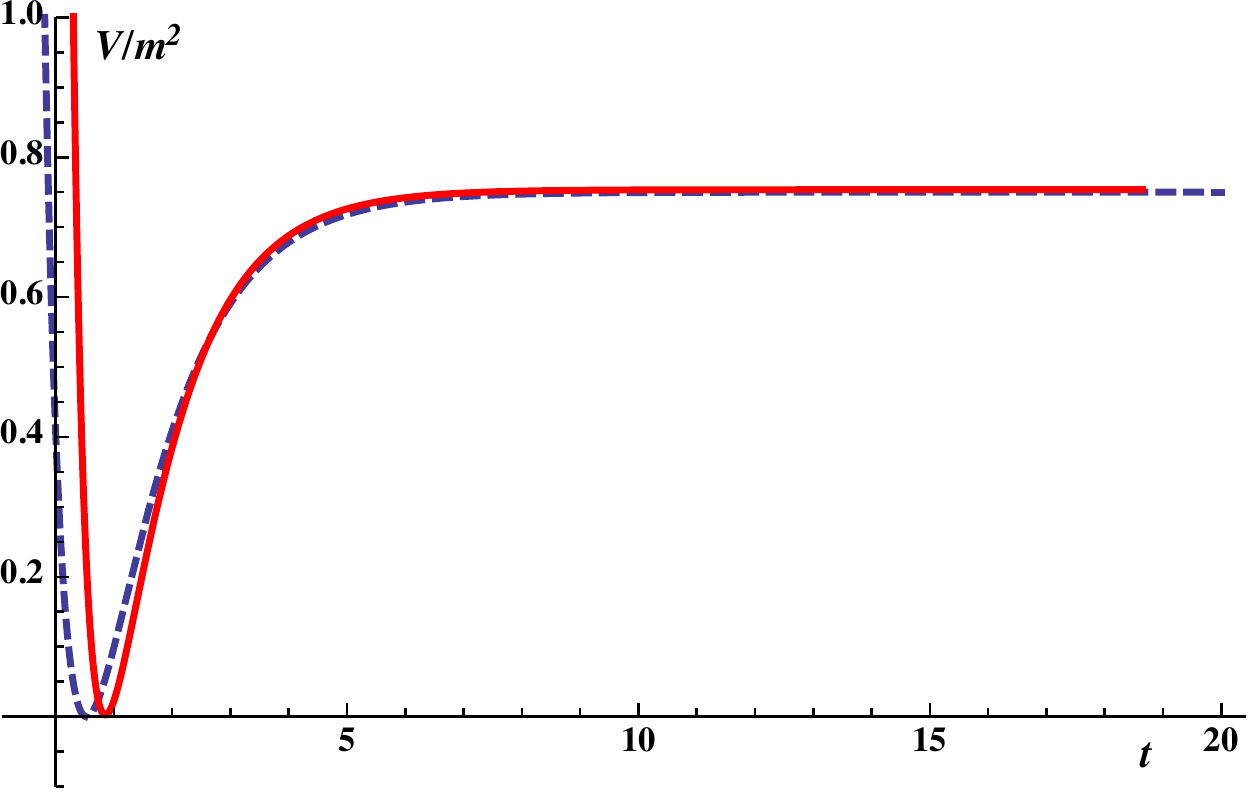}} 
	\caption{\it Projections of the effective inflationary potential for the model (\ref{tph_w}) with the
	Polonyi sector ((\ref{mu_pol_2}) and (\ref{Kpol2})), with $p=0$. 
	We use the nominal values $\Lambda_z=10^{-3}$ with ${\tilde m}^2/m = 0.9$. The values $\langle z\rangle$, $b$, and $\langle \phi\rangle$ are given approximately by (\ref{shift}) (shown by the dashed curve), and by (\ref{shift2}) (shown by the solid curve).
	} 
	\label{largemu}
\end{figure}

Unlike the WZ case discussed above, the inflationary potential maintains its form even for large ${\tilde m}$, and 
arbitrarily small $\Lambda_z$.  For example, in Fig.\,\ref{largemu}, we show the inflationary potential
with ${\tilde m}^2= 0.9 m M_P$ and $\Lambda_z = 10^{-3} M_P$ both the twisted (solid) and untwisted (dashed) Polonyi models. 
Here $t = \sqrt{3/2}\ln (2T)$ is the canonically normalized inflaton. In the figure, $\langle z\rangle$ and $b$ have been fixed at the approximate values given in (\ref{shift}) and (\ref{shift2}), respectively (higher order terms in (\ref{shift}) cannot be neglected).  

The gravitino mass for small $\Lambda_z$ can be written as
\begin{eqnarray}
m_{3/2} = m \frac{5 {\tilde m}^6 + 6 {\tilde m}^2 m^2 M_P^2}{2(3 m^2 M_P^2 + {\tilde m}^4)^{3/2}} \qquad {\rm untwisted}\,, \\
m_{3/2} = m \frac{ 4{\tilde m}^6  + 2 {\tilde m}^2 m^3 M_P}{2\sqrt{3}(m^2 M_P^2 + {\tilde m}^4)^{3/2}} \qquad {\rm twisted}\,,
\end{eqnarray}
which in the limit of small ${\tilde m}^2/m M_P$ for both cases gives the expected result $m_{3/2} = {\tilde m}^2/\sqrt{3}M_P$.
Indeed, for ${\tilde m}^2 = 0.9 m M_P$, we obtain $m_{3/2}\sim 10^{13}$ GeV. 
It is indeed rather surprising that even for a large Polonyi mass scale,
the inflationary dynamics are little affected. This is only true for case C given by Eq. (\ref{tph_w}).
Thus we are free to make the `natural' choice of ${\tilde m} = m \approx 10^{-5} M_P$.
In this case, the gravitino mass is
\beq
m_{3/2} = \frac{m^2}{\sqrt{3}M_P} \approx 0.2~{\rm EeV} .
\eeq
Furthermore, it was shown in the first reference of \cite{DW} that one also needs to impose $m_{3/2} < H$ in order to keep perturbative control of the K\"ahler potential.
In what follows, we will restrict our attention to the twisted Polonyi model.

\begin{figure}[h!]
\centering
\vskip -1.5in
	\scalebox{0.5}{ \hskip -.8in\includegraphics{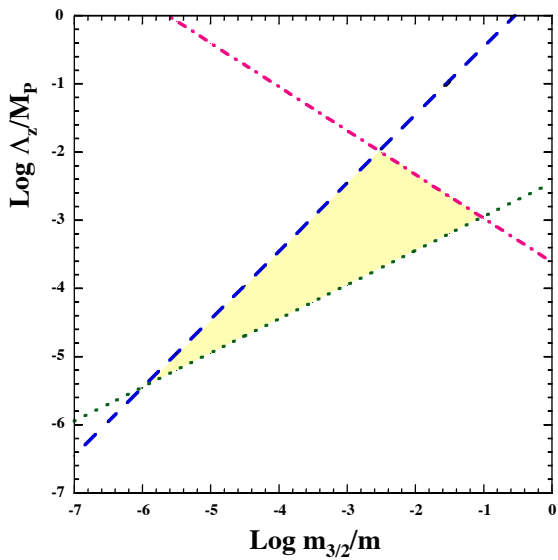}} 
	\vskip -1in
	\caption{\it Bounds on the stabilization parameter $\Lambda_z$ 
	as a function of the gravitino mass.
	The blue dashed line shows the upper limit on $\Lambda_z$ from the requirement that the mass of the Polonyi field
    lies above the inflaton mass.  The pink dot-dashed line (negatively sloped) is an upper limit on $\Lambda_z$ derived from
    an upper limit on the branching ratio of inflaton decays to gravitinos. 
    The green dotted line shows the lower limit on $\Lambda_z$,
    assuming $F < \Lambda_z^2$.
    The shaded region is allowed by all constraints. Note, however, that there are lower bounds on the gravitino mass given in Eqs. (\ref{ps4}) and (\ref{ps6}).
	} 
	\label{lambdalim}
	
\end{figure}

To avoid the production of the Polonyi field during reheating, we can derive
an upper limit on $\Lambda_z$ 
from the requirement that $m_z = \sqrt{12} m_{3/2}M_P/\Lambda_z > m$.
This limit is shown by the blue dashed line in Fig. \ref{lambdalim}.
Another upper limit on $\Lambda_z$ is obtained by requiring that the branching ratio of inflaton
decays to gravitinos does not lead to an excess abundance of gravitinos (discussed in more detail in section III.D).
This constraint is shown by the negatively sloped pink dot-dashed line.
 Acceptable parameters
lie below both lines (blue dashed and pink dot-dashed) 
Finally, we also have a lower bound on $\Lambda_z$,
stemming from our effective correction to the K\"ahler potential which imposes
 $F < \langle z\rangle < \Lambda_z^2$ or $1/2 \log (m_{3/2}/M_P) < \log (\Lambda_z/M_P)$. This lower bound is shown by the 
green dotted line in Fig. \ref{lambdalim}, and all values of $10^{-5} < m_{3/2}/m < 10^{-1}$ are allowed so long as 
 $\Lambda_z$ lies in the pale shaded region.

\section{The particle spectrum}

\subsection{Gaugino and Scalar masses}

We next discuss ways to generate gaugino and scalar masses via perturbative mediations of supersymmetry breaking, and then we will turn to more strongly-coupled mediation mechanisms.

One perturbative possibility for mediation of supersymmetry breaking is via gaugino mediation~\cite{gaugino_mediation}. New physics at a messenger mass scale $M$ generates
a coupling of the supersymmetry breaking field, $Z= z +\theta^2 F$ (ignoring fermionic components) to the gauge fields
\beq
\int d^2 \theta \, \left(f_0 + f_1 \frac{Z}{M} \right) {\rm tr}  W^\alpha W_\alpha~, \label{ps1} 
\eeq
where at leading order $f_0 = 1/(4g^2)$. This will give rise to gaugino masses of order
\beq
M_{1/2} \sim g^2 f_1 \frac{F}{M}~.  \label{ps2} 
\eeq

Scalar masses are then generated by SM loop corrections and will be of order
\beq
m_{0}^2 \sim  \frac{g^2}{16 \pi^2}  M_{1/2}^2~,  \label{ps3} 
\eeq
where $g$ is a SM gauge coupling. 
Usually there is a log enhancement of the scalar masses due to the running between the mediation scale $M$ and the scale of the superpartners.
In our case however, since the superpartners are very heavy there is not much running, so numerically we can use as an order of magnitude estimate
$m_0 \sim (\alpha/g)\,M_{1/2}\sim 0.06 M_{1/2}$. In order for the effective field theory to be well-defined (higher-dimensional operators are negligible), one needs to impose $F< M^2$. We also forbid inflaton decays into superpartners, which roughly requires $m_0 > 3 \times 10^{13}$ GeV.  Under such assumptions with $\sqrt{F}\sim M$,
one obtains $\sqrt{F} > 5 \times 10^{14}$ GeV and therefore the minimum value of the gravitino mass in such a scenario  is
\beq
m_{3/2}  > 6 \times 10^{10}\, {\rm GeV} = 60 \ {\rm EeV}~. \label{ps4} 
\eeq
It seems difficult to decrease the gravitino mass much below $100$ EeV with known perturbative mediations of supersymmetry breaking. For example, in a model based on gauge mediation, scalar masses would be expected to be of order the gaugino masses such that $m_0 \sim M_{1/2} \sim (g^2/16\pi^2) F/M$ (hence there is no relative loop suppression). The lower limit on $m_{3/2}$ is then increased to 
$
m_{3/2}  > 2\times 10^{13} \ {\rm GeV}~, \label{ps5} 
$
which is essentially incompatible with the required upper limit, $m_{3/2} < m$. 

In order to decrease the viable values of the gravitino mass, some strong coupling effects seem to be needed, like for example in holographic models of supersymmetry breaking of the type described in \cite{warped},  or general gauge mediation \cite{ggm}. In more generic terms,  this means a mediation mechanism with no loop suppression in the generation of visible sector soft masses, so that we obtain $m_0 \sim M_{1/2} \sim F/M$.
Thus we can start with the same gauge kinetic function as in (\ref{ps1}), namely
\beq
f_{\alpha\beta} = \left(f_0 + f_1 \frac{Z}{M}\right) \delta_{\alpha\beta} \, ,
\label{zkin}
\eeq
and generate squark/slepton/Higgs soft masses through operators of the type
\begin{equation}
\int d^4\theta~\frac{Z^\dagger Z}{M^2} Q^\dagger Q  \ , 
\end{equation}
where $Q$ denotes a generic MSSM chiral superfield.  
In this case, we recover the limit 
\beq
m_{3/2}  > 0.2 \ {\rm EeV}~, \label{ps6} 
\eeq
and corresponds to the bound derived in \cite{DMO}.
Perturbativity of the correction to gauge couplings $f_1 \frac{\langle z\rangle}{M} < f_0$ , together with the requirement, $M_{1/2} > m$
leads to the new lower limit
\begin{equation}
m_{3/2} > \frac{4\langle z\rangle}{\sqrt{3}}\frac{m}{M_P}~.
\end{equation}
However, this constraint is easily satisfied, once the other constraints, such as $F \leq M^2$ are taken into account.

Note that at the lower gravitino mass limit (\ref{ps6}) we obtain the bound $g^2 f_1\gtrsim 1$.
Since mediation is strongly coupled, this is not really surprising.  In both holographic models and general gauge mediation setups, additional states of mass, $M$ and heavier are expected. In order to not perturb our single-field inflation framework, the masses
of these states should be above the Hubble scale $H$ during inflation\footnote{It is also possible that all additional scalars obtain Hubble scale masses during inflation, therefore avoiding this condition.}, which implies  generically $M > H $. This condition is satisfied by the range of gravitino masses in Fig.~\ref{lambdalim} and is saturated at the lower bound. 

Finally, we comment on the partial wave unitarity limit arising from the scattering of two gluons into two gravitinos~\cite{Bhattacharya:1988tw}. For gaugino mediation, tree-level unitarity is violated at a scale $\simeq 17/(g^2 f_1) M$, which for 
$g^2 f_1 \lesssim 17$ is above the messenger scale $M$ (where new degrees of freedom should appear), and therefore compatible with the constraint arising from the gravitino mass limit (\ref{ps6}).

\subsection{Constraints on the scale of supersymmetry breaking from reheating}

Reheating proceeds by coupling the inflaton to the MSSM sector. Since all superpartners are above the inflaton mass and reheating temperature, reheating produces predominantly SM particles (the abundance of gravitinos is discussed in section III.D). Radiative corrections with MSSM fields in loops correct the inflaton potential. In low-energy supersymmetry such corrections are tiny, since they are proportional to the scale of supersymmetry breaking \cite{ENOT}. In our case with high-scale supersymmetry breaking,  there may be large radiative corrections that can spoil flatness of the inflaton potential. Such constraints can put upper limits on the superpartner masses and therefore
on the gravitino mass. 

For example, a direct coupling (through the gauge kinetic function)
of the inflaton, $t$ to gauge fields, $f \ni h_1 t/M_P$, would induce
quadratic and quartic corrections of magnitude
\begin{eqnarray}
\label{qqcor}
\delta m^2 \sim \frac{h_1^2}{16 \pi^2}M_{1/2}^2 \,,  \\
\delta \lambda \sim \frac{h_1^4}{16 \pi^2}\frac{M_{1/2}^2}{M_P^2}\,,
\label{qqcor1}
\end{eqnarray}
which both place non-trivial bounds on the coupling $h_1$.
For reheating dominated decays to gauge bosons, this can be translated 
into a limit on the reheating temperature and eventually the 
gravitino abundance. 

As we discuss in more detail in section III.D, reheating in this model proceeds
via the gravitational coupling of the inflaton to two Higgs bosons. 
The coupling of the inflaton field, $t$ to MSSM fields was derived in \cite{EGNO4},
and the relevant bosonic coupling is 
\begin{eqnarray}
&& \mathcal{L}_{\rm eff} \ni \frac{{\rm Re} T }{\sqrt{3}} (n_I + n_L - 3) W^{IL}{\bar W}_{LJ} \Phi_I {\bar \Phi}^J\,, \nonumber \\
&& \sim \mu^2 e^{\sqrt{\frac{2}{3}} t} (|h_u|^2 + |h_d|^2 )  \ , 
\label{higgscoup}
\end{eqnarray}
where $n_{I,J}$ are modular weights\footnote{Note that the  definition of modular weights in our paper is opposite in sign with respect to the standard convention.} of the superfields $\Phi_{I,J}$
and should be taken to be equal to one for untwisted Higgs fields. 
The coupling for the Higgs fields is then $\mu^2/\sqrt{3}M_P$, where $\mu$
is the MSSM Higgs mixing mass, which is now expected to be of order the
scalar masses. The quadratic and quartic corrections in Eqs.~(\ref{qqcor})-(\ref{qqcor1})
are found with the replacement $h_1 \to \mu^2/mM_P$. 
Requiring $\delta m^2 \ll m^2$ and $\delta \lambda \ll 10^{-14 }$
sets a rough bound on $\mu/m \lesssim 10^2$ which we will see
below is satisfied when $\mu$ is adjusted to give the correct 
gravitino relic density. However a  more model-independent statement is that reheating sets constraints on inflaton couplings to MSSM fields. Once we fix such couplings, the requirement that quantum corrections do not spoil flatness of the inflaton potential generically set upper bounds on superpartner masses.    
Such bounds are dependent on the inflationary model and details of reheating, but they typically indicate that the scale of supersymmetry breaking should not be too much higher than the inflationary mass scale.

\subsection{The Higgs mass and vacuum stability}

The fact that the sfermion, gaugino and Higgsino masses are above $3\times 10^{13}$~GeV, leads to important implications for the Higgs boson mass and vacuum stability. It is well-known that to obtain the
125 GeV Higgs mass for $\tan\beta\lesssim 50$, the maximum supersymmetry breaking scale is approximately $10^{10}$~GeV. However in Ref.~\cite{susyhd}, it was noted that the supersymmetry breaking scale can be increased to the GUT scale ($\sim 10^{16}$~GeV), for very large values of $\tan\beta\sim 200$. This assumes a degenerate superpartner spectrum (at $\widetilde m$), with a bottom superpotential Yukawa coupling, ${\widehat y}_b =y_b/\cos\beta$ such that ${\widehat\alpha}_b= {\widehat y}_b^2/(4\pi)\sim 0.5$. Even though this coupling is perturbative, a very close Landau pole develops at $\Lambda\sim 10 {\widetilde m}$.

\begin{figure}[h!]
\centering
	\scalebox{0.45}{ \hskip -.8in\includegraphics{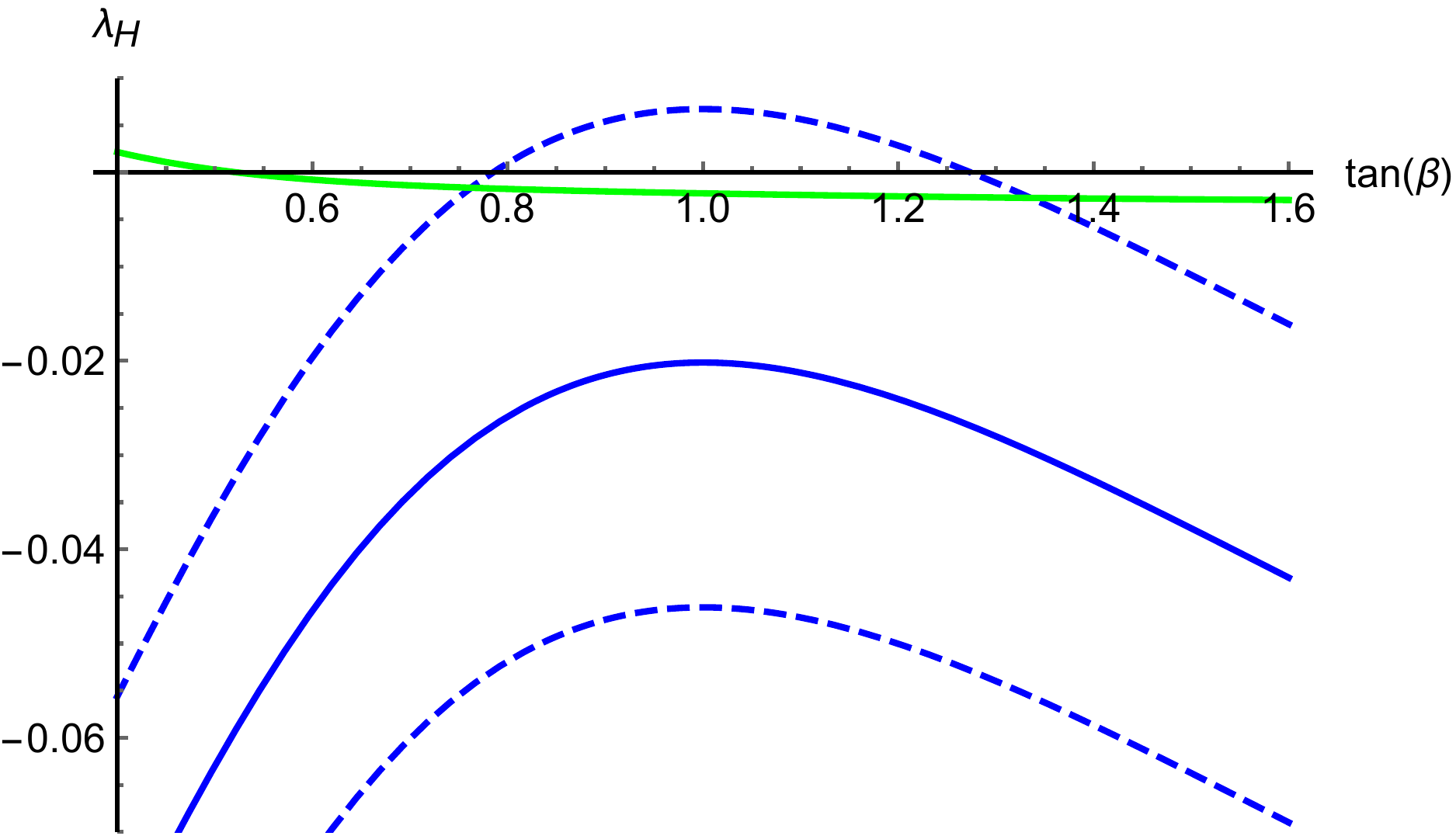}} 
	\caption{\it The Higgs quartic coupling, $\lambda_H$ at the matching scale ${\tilde m}=5\times 10^{13}~{\rm GeV}$, as a function of $\tan\beta$.  The solid blue line is the difference between the SM and tree-level SUSY values, with the dashed lines indicating the $\pm 3\sigma$ contours due to the error in the top-quark Yukawa coupling $y_t(m_t)$, where $m_t= 173.1 \pm 0.7~{\rm GeV}$~\cite{Degrassi:2012ry}.
The green line shows the one-loop threshold correction arising from our superpartner spectrum.
	} 
	\label{fig:lamH}
\end{figure}

Instead, Ref.~\cite{ellisjr} considered non-degenerate superpartner masses, and showed that GUT scale masses can in fact be accommodated for much smaller values of $\tan\beta$. In particular, assuming ${\widetilde m}=m_{\tilde Q_L,3}=m_{{\tilde t}_R}=10^{16}$~GeV and $\tan\beta=1$, a scan of gaugino and 
first/second generation sfermion masses in the range $[{\widetilde m}, 100 {\widetilde m}]$, and Higgsino masses in the range $[\frac{{\widetilde m}}{100}, {\widetilde m}]$ gives rise to the required threshold corrections of the Higgs quartic coupling.

The analysis in Ref.~\cite{ellisjr} suggests that our superpartner spectrum can give rise to similar threshold corrections needed for a 125 GeV Higgs mass. For concreteness we will consider a gaugino-mediated spectrum, and expect similar qualitative features for a spectrum generated by a more strongly-coupled mediation mechanism. 
Identifying the right-handed stau mass with the scale ${\widetilde m} = m_{\tilde \tau_{R}}\sim 5\times 10^{13}$~GeV, will then determine the size of the remaining sfermion and gaugino masses. Assuming that the gaugino masses are generated at a scale $m<\mu_0 \leq 10^{16}$~GeV, the sfermion masses are then approximately given by
\begin{eqnarray}
m_{\tilde Q}^2 &\simeq& \frac{1}{16\pi^2}\left(\frac{32}{3} g_3^2+6 g_2^2 +\frac{2}{15}g_1^2 \right) M_{1/2}^2\log\frac{\mu_0}{\tilde m}\,,\\
m_{\tilde u}^2 &\simeq& \frac{1}{16\pi^2}\left(\frac{32}{3} g_3^2+\frac{32}{15}g_1^2 \right) M_{1/2}^2\log\frac{\mu_0}{\tilde m}\,,\\
m_{\tilde d}^2 &\simeq& \frac{1}{16\pi^2}\left(\frac{32}{3} g_3^2+\frac{8}{15}g_1^2 \right)M_{1/2}^2\log\frac{\mu_0}{\tilde m}\,,\\
m_{\tilde L}^2 &\simeq& \frac{1}{16\pi^2}\left(6 g_2^2 +\frac{6}{5}g_1^2 \right) M_{1/2}^2\log\frac{\mu_0}{\tilde m}\,,\\
m_{\tilde e}^2 &\simeq& \frac{1}{16\pi^2}\left(\frac{24}{5}g_1^2 \right) M_{1/2}^2\log\frac{\mu_0}{\tilde m}\,.
\end{eqnarray}
Other soft parameters such as $A_t$, $B\mu$ and $m_{H_{u,d}}^2$ are also generated radiatively. Using this approximate spectrum we can then compute the one-loop threshold corrections as given
in Ref.~\cite{Bagnaschi:2014rsa,ellisjr}. The result in shown in Figure~\ref{fig:lamH}. The contribution to the Higgs quartic coupling from our superpartner spectrum overlaps with the $\pm3 \sigma$ band of the $\lambda_H$ coupling provided that $0.75\lesssim \tan\beta \lesssim 1.34$.

However, note that the one-loop threshold correction matches at a ${\it negative}$ value of the Higgs quartic coupling. This occurs because of our somewhat compressed spectrum,
$m_{\tilde Q}/{\tilde m}\sim 1.9, M_{1/2}/{\tilde m}\sim 2, \mu/{\tilde m}\sim 1.5$, and the fact that the gaugino/Higgsinos contribute negatively,
while the positive stop contribution is suppressed by mixing near $\tan\beta\sim 1$. 
A more detailed determination of the superpartner spectrum that gives different mass ratios may lead to a positive correction and then the matching could occur for positive values of $\lambda_H$.

Nevertheless even with a positive one-loop threshold correction from our superpartner spectrum, the value of the top quark Yukawa coupling, $y_t(m_t)$ would need to be near its $-3\sigma$ extreme value that is allowed by the large uncertainty in the top-quark mass measurement. Otherwise one needs to rely on other threshold effects
to stabilize the Higgs potential below the scale $\tilde m$. 
For instance this could be due to the inflaton coupling to the Higgs, $y_I \sim 
\frac{1}{\sqrt{3}} \frac{\mu^2}{m M_p}\sim 10^{-3}$. However, assuming that the supersymmetric and soft contributions to the inflaton mass are of the same order, this coupling causes a shift~\cite{Giudice:2011cg} 
in the Higgs quartic coupling by an amount
given by $\delta\lambda_H \simeq y_I^2 \sin^22\beta \leq y_I^2\sim 10^{-6}$, 
which is negligibly small.

Alternatively, a heavy scalar singlet with a quartic coupling to the Higgs, could be introduced that is related to the generation of the neutrino masses~\cite{EliasMiro:2012ay}. If this correction causes the Higgs quartic to be large and positive at the SUSY scale ${\tilde m}$, then a negative one-loop threshold correction would not be a problem, since it could then be absorbed by the SUSY tree-level contribution for sufficiently large $\tan\beta$. Nonetheless this does require an extra tuning in the model in order that this scalar singlet remains light (at an intermediate scale).

Another potential concern is that the 
required $\tan\beta$ values in Figure~\ref{fig:lamH} are near one, and normally in the MSSM this would cause a Landau pole in the top quark Yukawa coupling to appear below the GUT scale. However since our sparticle spectrum is quite heavy, the top quark Yukawa coupling is reduced by a factor of two at the scale $\widetilde m\simeq 5\times 10^{13}$~GeV. The matching condition for the top Yukawa coupling $y_t ={\widehat y}_t \sin\beta$, where ${\widehat y}_t$ is the superpotential Yukawa coupling, then allows for a lower value of $\tan\beta$, with a corresponding larger value of ${\widehat y}_t$. Since there is relatively little running in our high-scale SUSY model above the scale ${\widetilde m}$, the larger ${\widehat y}_t$ value can remain perturbative below the GUT scale.

Furthermore, to radiatively break electroweak symmetry requires that at some scale, $\frac{d m_{H_u}^2}{dt} = 0$, or ${\widehat y}_t m_{\tilde Q} \sim g_2 M_2$. In the gaugino-mediated model this condition occurs when $\tan\beta \sim 0.5$, which is incompatible with the range required in Figure~\ref{fig:lamH}~\footnote{In addition, requiring that there is no color-breaking minimum deeper than the electroweak minimum~\cite{Bagnaschi:2014rsa}, leads to $\tan\beta\gtrsim 0.6$ for our spectrum.}. However if there were a new positive contribution to the Higgs quartic coupling then it may be possible to also achieve radiative electroweak symmetry breaking. For instance, starting the running above the GUT scale $10^{16}$ GeV would give different sfermion mass ratios to make $\delta\lambda_H$ positive and/or allow $m_{H_u}^2$ to run negative before ${\widetilde m}$. The details of these possibilities are beyond the scope of this paper and will be left for future work.

\vspace{3cm}

\subsection{Dark Matter}

One of the main motivations of our high-scale SUSY model is its ability to account for the 
dark matter in the form of gravitinos with masses $m_{3/2} \gtrsim 0.2$ EeV.
Because the supersymmetric particle spectrum lies above the inflaton mass,
the dominant mechanism for gravitino production becomes 
SM + SM $\to$ 2 gravitinos with longitudinal polarizations \cite{bcdm,DMO}
or the decay of the inflaton directly to gravitinos depending on the 
reheating temperature.

The gravitino production rate was derived in 
\cite{bcdm}
\beq
R = n^2 \langle \sigma v \rangle \simeq 21.65 \times \frac{T^{12}}{F^4}\,,
\label{Eq:r}
\eeq
where $n$ is the number density of incoming states.
This temperature dependence can be understood as follows:  one uses
$n \propto T^3$, and we expect the gravitino production cross section to scale as  $\langle \sigma v \rangle
\propto T^6/F^4$. 
From the rate $R(T)$, we can determine that $\Gamma \sim R/n \sim T^9/M_P^4 m_{3/2}^4$ (assuming $m_{3/2} \ll {\widetilde m} $)
leading to a gravitino abundance $n_{3/2}/n_\gamma \sim \Gamma/H \sim T^7/M_P^3 m_{3/2}^4$ evaluated at $T=T_{RH}$ or 
\beq
\Omega_{3/2}h^2 
\simeq
0.11 \left( \frac{0.1 ~\mathrm{EeV}}{m_{3/2}} \right)^3
\left( \frac{T_{RH}}{2.0 \times 10^{10}~\mathrm{GeV}} \right)^7\,,
\label{Eq:omega}
\eeq
assuming instantaneous decay and thermalization. 
Thus, thermal production of gravitinos with $m_{3/2} > 0.2$ EeV
would require $T_{RH} > 3 \times 10^{10}$ GeV.

It is known however, that the reheating process is not instantaneous,
and that the temperature of the Universe during inflaton decay
can exceed $T_{RH}$ by orders of magnitude \cite{Tmax,egnop} up to a value $T_{max}$.
Due to the strong temperature dependence of the gravitino production 
cross section, there will be significant production of gravitinos at $T_{max}$,
which is not fully diluted by the entropy produced in subsequent decays. 
The final gravitino abundance in this case (with $\sigma \propto T^6$) 
relative to the instantaneous approximation is \cite{gmop}
\beq
r_{3/2} = \frac{56}{5}\ln \left(\frac{T_{max}}{T_{RH}}\right)\,,
\label{enhance}
\eeq
where 
\beq
T_{\rm max} \simeq 0.5\left(\frac{m}{\Gamma_T}\right)^{1/4}T_{RH}\,, 
\eeq
for inflationary models of this type, where $\Gamma_T$ is the total inflaton decay rate. 

For the inflationary model discussed above, there are many possible decay 
channels all of which are Planck suppressed. The decay channel given in 
Eq. (\ref{higgscoup}) is ordinarily (with weak scale supersymmetry breaking) 
negligible as $\mu^2/(m M_P) \ll 1$. However, in our case, since $\mu > m$,
this is actually the dominant inflaton decay mode $t\to H_{u,d} {H^*}^{d,u}$
which ultimately corresponds to a decay of $t \to h h$ where 
$h$ is the SM Higgs boson. The decay rate to two Higgs bosons is \cite{EGNO4}
\beq
\Gamma_{2h} = \frac{\mu^4}{384\pi m M_P^2}\sin^2 2\beta\,,
\eeq
where an additional factor of 1/16 has been included in writing 
$H_u^0 = h/\sqrt{2} \sin \alpha$, and $H_d^0 = h/\sqrt{2} \cos \alpha$,
and noting that $\alpha = \beta$ in the high scale SUSY limit.

If we define an effective Yukawa-like coupling, $y_I = \mu^2/(4\sqrt{3}m M_P)$,
such that $\Gamma_{2h} = \frac{y_I^2}{8\pi}m$, we can express the reheating temperature
in terms of $y_I$ \cite{ps2,egnop,DMO} 
\beq
T_{RH} = \left(\frac{10}{g_s} \right)^{1/4} \left(\frac{  2 \Gamma_{2h}\,M_P}{\pi\,c} \right)^{1/2} = 0.5 \frac{y_I}{2\pi} \left( m\,M_P \right)^{1/2} \, ,
\label{Eq:trh}
\eeq
where $g_s$ is the effective number of light degrees
of freedom, in this case set by the Standard Model, $g_s = 427/4$ and 
$c\approx 1.2$
is a constant. We can then re-express the relic abundance (\ref{Eq:omega}) as
\bea
\Omega_{3/2} h^2 \simeq 0.11\, r_{3/2} &&\left( \frac{0.1 ~\mrm{EeV}}{m_{3/2}} \right)^3
\left( \frac{m}{3 \times 10^{13}\, {\rm GeV}} \right)^{7/2} \nonumber\\
&&\times \left( \frac{y_I}{2.9 \times 10^{-5}} \right)^7\,,\nonumber\\
= 0.11\, r_{3/2} &&\left( \frac{0.1 ~\mrm{EeV}}{m_{3/2}} \right)^3
\left( \frac{3 \times 10^{13}\, {\rm GeV}}{m} \right)^{7/2} \nonumber\\
&&\times\left( \frac{\mu}{ 1.2\times 10^{14}\, {\rm GeV}} \right)^{14}\,,
\label{Eq:omegamu}
\eea
where we have included the enhancement factor $r_{3/2}$ from Eq. (\ref{enhance}). The enhancement factor depends on $\ln \mu$, and for the range
of $\mu$ values considered here (roughly $10^{14} - 10^{15}$ GeV), $r_{3/2}$
varies very little and we take it as a constant $r_{3/2}=25$.

\begin{figure}[h!]
\centering
\vskip -1.3in
	\scalebox{0.5}{ \hskip -.8in\includegraphics{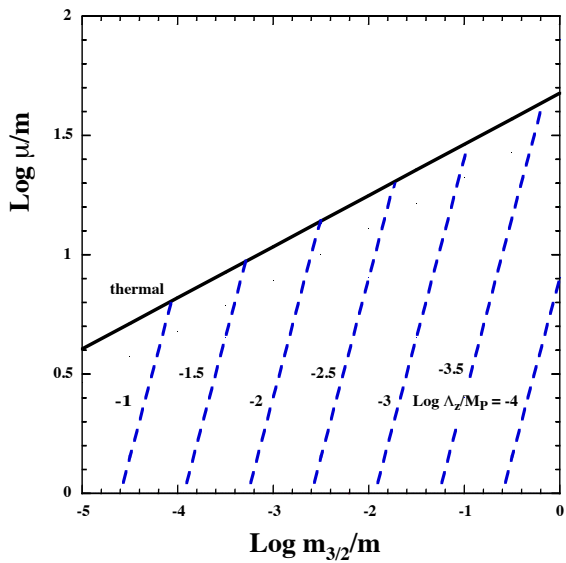}} 
	\vskip -1in
	\caption{\it The value of $\mu$ relative to the inflaton mass needed to obtain
    the correct relic density of gravitinos thermally through reheating (solid line) as a function of the gravitino mass.
    Also shown (dashed lines) are the values of $\mu$ needed to obtain
    the correct relic density of gravitinos through inflaton decays
    for a given value of $\log \Lambda_z/M_P$ as labelled.
    }
	\label{b32mu}
\end{figure}

The value of $\mu$ needed to obtain the correct relic density of gravitinos is
shown by the solid line in Fig. \ref{b32mu} using Eq. (\ref{Eq:omegamu}). It is rather amazing that independent of the supersymmetric
particle spectrum discussed above, the value of $\mu$ needed for the 
correct abundance of gravitinos is in the range of roughly 3-30 times the 
inflaton mass. This is exactly where one might 
expect the Higgsino mass to lie
given our spectrum of heavy scalars and gauginos.

It is also possible that $\mu$ takes values below the solid line in Fig. \ref{b32mu}. In that case, the abundance of gravitinos is below the 
needed relic density of dark matter (by the same token, values of $
\mu$ above the solid line are excluded as they yield a relic density in 
excess of the observed one). Nevertheless, it is still possible
to recover the correct relic density through inflaton decay to gravitinos. 
The gravitino abundance produced by inflaton decay for a given branching fraction 
to gravitinos, $B_{3/2} = \Gamma_{3/2}/\Gamma_{2h}$, was computed in \cite{DMO}
\bea
\Omega_{3/2}^{decay}h^2 = 0.11
&&\left(\frac{B_{3/2}}{1.3\times 10^{-13}}\right) \left(\frac{y_I}{2.9\times 10^{-5} } \right) 
\\
&&
\times
\left(\frac{m_{3/2}}{0.1~{\rm EeV}} \right)  \left(\frac{3 \times 10^{13} ~{\rm GeV}}{m} \right)^{1/2}.
\nonumber
\label{Ob32}
\eea
The decay of the inflaton to two gravitinos was computed in \cite{EGNO4} with 
\beq
\Gamma_{3/2} = \left(\frac{\Lambda_z}{M_P} \right)^4 \frac{3 m_{3/2}^2 m}{256\pi M_P^2}\,,
\eeq
so that
\beq
B_{3/2} = \frac{9}{2}\left(\frac{\Lambda_z}{M_P} \right)^4 \left(\frac{m_{3/2}}{m}\right)^2 \left(\frac{m}{\mu} \right)^4\,.
\eeq
Using Eq. (\ref{Ob32}), inputting $B_{3/2}$ and $y_I$,
we obtain the correct relic density of gravitinos along the sloped
dashed and dotted lines for different values of $\Lambda_z/M_P$ as labelled.

As one can see from Fig.\ref{b32mu}, the value of the abundance of gravitinos from the decay of the inflaton is 
strongly dependent on the value of $\Lambda_z$ as this scale controls the branching ratio $B_{3/2}$. 
As a consequence, we can derive an upper limit to $\Lambda_z$
\beq
\frac{\Lambda_z}{M_P} \le 2.4 \times 10^{-4} \left( \frac{m}{m_{3/2}} \right)^{9/14}~,
\eeq
which is shown in Fig. \ref{lambdalim} by the negatively sloped pink dot-dashed line.
This upper limit in turn imposes an upper limit to the gravitino mass $m_{3/2} \lesssim 0.1 m$.

\section{Conclusions}

It may be that supersymmetry is not physically realized at energy scales accessible to the LHC.
The hierarchy problem and naturalness biased our expectations that the supersymmetry mass scale
was at or near the weak scale making experimental discovery all but inevitable.
Such is not (yet) the case, and the mass scale of the supersymmetric spectrum remains unknown.
It is therefore plausible to consider the possibility that nearly the entire supersymmetric spectrum
lies at very high energies.  If it is above the inflationary scale it is quite possible that the rich spectrum 
of supersymmetric partners were never produced in the early universe after inflationary reheating.

The exception could be the gravitino whose mass may remain below the inflaton mass. 
In this case, gravitinos could be produced (in pairs) during reheating \cite{bcdm,DMO}
and because of the strong sensitivity to temperature, have enhanced production
at the start of reheating when the temperature of the radiation plasma is above the reheating scale \cite{gmop}.
For a sufficiently high reheating temperature ($T_{RH} \gtrsim 10^{10}$ GeV), thermally produced 
gravitinos would have the correct relic density to account for the observed cold dark matter in the Universe.

In this paper, we have constructed a working model incorporating both inflation and supersymmetry breaking
which leads to the heavy supersymmetric spectrum with gravitino dark matter. Our starting point is 
no-scale supergravity \cite{no-scale,LN}. In the family of models formulated in Ref. \cite{Cecotti},
the inflaton is associated with the volume modulus, $T$, and the scalar potential is identical to that
derived in the Starobinsky model \cite{Staro}. Supersymmetry breaking is achieved by adding a 
strongly stabilized Polonyi field which preserves the potential for inflation. Reheating occurs through the 
gravitational coupling of the inflaton to the Standard Model Higgs scalars. Because
supersymmetry breaking occurs at a high scale, the $\mu$ parameter is large (larger than the inflaton mass)
and the dominant decay channel is to two Higgs bosons. Depending on the value of $\mu$, gravitinos may
be produced through reheating with the correct relic density over a wide range of 
gravitino masses. For smaller values of $\mu$ (but still larger than the inflaton mass),
the correct relic density may be obtained via the decays of the inflaton directly to gravitinos. All constraints are satisfied for gravitino masses in the range 
$0.1~{\rm EeV} \lesssim m_{3/2} \lesssim 1000~{\rm EeV}$.

While we have shown that the Higgs mass in this class of models can be compatible with the
experimental value when $\tan \beta \approx 1$, we leave open for future study
the questions of vacuum stability, and radiative electroweak symmetry breaking - two attractive 
features normally associated with supersymmetric models.  Both stability and symmetry breaking 
can be achieved without supersymmetry via an intermediate scale such as in SO(10) grand unification~\cite{mnoz}, and therefore suggests that in models of high scale supersymmetry breaking, grand unification plays a crucial role in determining the Higgs mass.

Finally, the minimal setup of our model predicts no signatures in either collider or direct/indirect dark matter searches. Instead, scalars with masses near the Hubble scale during inflation could lead to non-gaussianities in the CMB that may eventually be observed. Alternatively, by introducing R-parity violation in the lepton-Higgs sector, the gravitino can become unstable with a suitably long-lived decay. The detection of the decay products, such as very high energy neutrinos or photons at HAWC or the Pierre Auger Observatory, would then be a possible sign of high scale supersymmetry with an EeV gravitino, and thus these types of signatures are worthy of further study.

\section*{Acknowledgements}
We would like to thank Sebastian Ellis for helpful discussions.
This  work was supported by the France-US PICS no. 06482.
 Y.M.  acknowledges partial support from the European Union FP7 ITN INVISIBLES (Marie
Curie Actions, PITN- GA-2011- 289442) and  the ERC advanced grants  
 Higgs@LHC. E.D. acknowledges partial support from the ANR Black-dS-String. The work of T.G. and  K.A.O. was supported in part
by the DOE grant DE--SC0011842 at the University of Minnesota.

\end{document}